\documentclass{article}

\usepackage{arxiv}

\usepackage[utf8]{inputenc} % allow utf-8 input
\usepackage[T1]{fontenc}    % use 8-bit T1 fonts
\usepackage{hyperref}       % hyperlinks
\usepackage{url}            % simple URL typesetting
\usepackage{booktabs}       % professional-quality tables
\usepackage{amsfonts}       % blackboard math symbols
\usepackage{nicefrac}       % compact symbols for 1/2, etc.
\usepackage{microtype}      % microtypography		% Can be removed after putting your text content
\usepackage{graphicx}
\usepackage{natbib}
\usepackage{doi}
\usepackage{amsmath,amssymb,amsfonts}
\usepackage{algorithmic}
\usepackage{textcomp}
\usepackage{subcaption}
\usepackage{wrapfig}

\title{FPCA: \underline{F}ield-Programmable \underline{P}ixel \underline{C}onvolutional \underline{A}rray for Extreme-Edge Intelligence}

\author{ \href{https://orcid.org/0000-0001-7216-8214}{\includegraphics[scale=0.06]{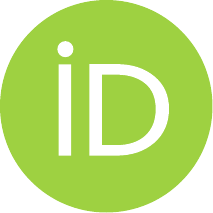}\hspace{1mm}Zihan Yin}\\
	Department of Electrical \& Computer Engineering\\
	University of Wisconsin-Madison\\
	Madison, WI 53704 \\
	\texttt{zyin83@wisc.edu} \\
	\And
	Akhilesh Jaiswal \\
	Department of Electrical \& Computer Engineering\\
	University of Wisconsin-Madison\\
	Madison, WI 53704 \\
	\texttt{akhilesh.jaiswal@wisc.edu} \\
}

\date{}

\renewcommand{\shorttitle}{FPCA}

\hypersetup{
pdftitle={FPCA: Field-Programmable Pixel Convolutional Array for Extreme-Edge Intelligence},
pdfsubject={Sensor},
pdfauthor={Zihan Yin, Akhilesh Jaiswal},
pdfkeywords={In-pixel Computing, Non-volatile Memory, In-sensor Computing, Reconfigurability.},
}

\begin{document}
\maketitle

\begin{abstract}
	The rapid advancement of neural network applications necessitates hardware that not only accelerates computation but also adapts efficiently to dynamic processing requirements. While processing-in-pixel has emerged as a promising solution to overcome the bottlenecks of traditional architectures at the extreme-edge, existing implementations face limitations in reconfigurability and scalability due to their static nature and inefficient area usage. Addressing these challenges, we present a novel architecture that significantly enhances the capabilities of processing-in-pixel for convolutional neural networks. Our design innovatively integrates non-volatile memory (NVM) with novel unit pixel circuit design, enabling dynamic reconfiguration of synaptic weights, kernel size, channel size and stride size. Thus offering unprecedented flexibility and adaptability. With using a separate die for pixel circuit and storing synaptic weights, our circuit achieves a substantial reduction in the required area per pixel thereby increasing the density and scalability of the pixel array. Simulation results demonstrate dot product operations of the circuit, the non-linearity of its analog output and a novel bucket-select curvefit model is proposed to capture it. This work not only addresses the limitations of current in-pixel computing approaches but also opens new avenues for developing more efficient, flexible, and scalable neural network hardware, paving the way for advanced AI applications.
\end{abstract}

% keywords can be removed

\section{Introduction}
\label{sec:introduction}
 Given the proliferation of high-resolution and high-frame rate imaging, the surge in data generated by cameras for artificial intelligence (AI) enabled computer vision (CV) applications presents significant energy and bandwidth challenges. This issue is exacerbated by the conventional architecture that separates the sensor from the processing units, leading to inefficiencies in data transfer and computation \cite{auto_driving, obj_track}. To address this, recent research has pivoted towards strategies that \textit{process and compress} data closer to the point of capture, aiming to alleviate these bottlenecks.
 
 To enhance efficiency in data processing and compression closer to the sensor, three primary methodologies have emerged, differentiated by how closely the processing unit is integrated with the data generation (sensor) unit:
\begin{enumerate}
    \item Near-Sensor Processing: In this setup, the data processor is located in close proximity to the CMOS image sensor (CIS) chip. This arrangement boosts energy and bandwidth efficiency by reducing the distance between the sensor and the processor \cite{near_sensor_sony, near_in_sensor_survey}. Despite this, the processor and sensor remain on separate chips, meaning there is still a notable distance including significant off-chip communication that data must traverse.
    \item In-Sensor Processing: This method incorporates either an analog or digital signal processor directly into the periphery of the sensor chip. By doing so, it significantly reduces the physical distance between where data is generated and first processed \cite{in_sensor_sleep_spotter}. This approach helps reduce the data transfer bottleneck between sensor and processor, yet the bottlenecks in transferring data from the sensor to the processor still exists as the sensor is physically separate from the processing unit that resides in the periphery outside sensor array.
    \item In-Pixel Processing: Taking integration of sensor and processor a step further, this innovative strategy equips each pixel circuit within the sensor with the ability to perform massively parallel computations. This means processing can start at the very site of data capture, along each pixel row(s) and/or column(s), drastically diminishing the bandwidth requirement and reducing both the energy consumption and latency of the sensor-processor system.
\end{enumerate}

Despite various efforts in in-pixel processing \cite{senputting, scamp_simd, prog_kernel, in_pixel_cifar10, aps_p2m, aps_p2m_detrack}, most of the works fail to accomplish high accuracy in complex machine learning (ML) tasks. Complex ML tasks often require advanced operations like multi-bit, multi-channel convolution, batch normalization (BN), and Rectified Linear Units (ReLU). Most approaches \cite{senputting, scamp_simd, prog_kernel, in_pixel_cifar10} have limitations, such as binary weight implementation and a lack of multi-channel convolution capability, while focusing on simpler datasets that do not fully represent the complexities of real-world CV applications. Notably, prior works \cite{aps_p2m, aps_p2m_detrack} have demonstrated significant improvements but are hindered by the fixed nature of the transistor-width-based weight implementation as well as the number of kernels and the stride size, which lacks the flexibility and reconfigurability needed for diverse CV applications. Thus, current research has failed to achieve the combination of high accuracy and reconfigurability for pixel sensors to tackle multiple CV applications within the same pixel array.
% This work introduces a novel approach that leverages a hybrid CMOS-NVM circuit and a reconfigurable structure, allowing for dynamic weight, channel, kernel reconfiguration to cater to various machine vision tasks. This advancement is pivotal, as it addresses the rigidity of previous systems by enabling adaptable weight settings post-fabrication, thus broadening the applicability of in-pixel processing solutions.

Furthermore, in-pixel processing involves co-design considerations across chip integration, circuit configurations, and algorithm. In previous attempts, embedding computational tasks like matrix-vector multiplication within the pixel array can compromise pixel density due to requirement of hundreds of added transistors per pixel based on size of stride and number of channels. Innovations such as heterogeneous 3D integration offer a pathway to vertically stack logic or memory substrates with the CIS\cite{yin2023design,kaiser2023technology}, potentially overcoming these limitations. However, the pixel pitch limits the number of transistors which in turn limits the size of stride, channels and kernel size of an neural network which effects the overall accuracy of the CV task.
% In this article, we are proposing a novel Field-programmable Camera Array(FPCA) pixel architecture design that aims to provide reconfigurability in weight value, kernel size, channel size and stride size. 
% Moreover, our approach significantly reduces the required area per pixel by eliminating the need for additional weight transistors and placing the additional weight block composed of non-volatile memory(NVM) in a separate weight die using 3D integration technology. This advancement not only enhances the density and scalability of the pixel array but also opens up new possibilities for deploying more complex neural network models directly onto the hardware. Through detailed simulation results, we demonstrate the effectiveness of our design in performing dot product operations— a fundamental building block of neural network computations— with improved efficiency and adaptability compared to existing designs.

This work introduces, for the first time, a novel Field-Programmable Pixel Convolutional Array (FPCA) pixel architecture design for in-pixel processing by enabling comprehensive field-programmability of all the key parameters within the initial layers of modern convolutional deep learning networks. Leveraging a hybrid CMOS-NVM circuit, this system introduces a reconfigurable structure designed to dynamically adjust all key parameters for state-of-the-art deep learning networks: 
\begin{enumerate}
    \item \textbf{Weight values}: allowing for modifications based on varying algorithm demands.
    \item \textbf{Channel configurations}: adapting to different data bandwidth and processing needs.
    \item \textbf{Kernel sizes}: customizable to match the specific convolutional requirements of different applications.
    \item \textbf{Stride sizes}: offering flexibility in feature sampling and data throughput.
\end{enumerate}

Such reconfigurability is crucial for creating \textit{sustainable solutions}, wherein a single hardware setup can meet a wide range of deep learning algorithm specifications. For instance, complex datasets like BDD100K\cite{bdd100k} might require smaller filter sizes and strides to capture details, while simpler datasets like  visual wake word (VWW)\cite{chowdhery2019visual}  can work with larger kernels and broader strides while maintaining classification accuracy. By incorporating these adjustable parameters, the FPCA not only addresses the static nature of previous systems but also significantly enhances the scope and application space for in-pixel processing use cases. This adaptability leads to field-programmability, ensuring that the same physical infrastructure can continuously evolve in response to emerging computational challenges and advancements in machine learning methodologies involving convolutional operations. Moreover, our approach significantly reduces the required area per pixel by eliminating the need for weight transistors within each pixel unit and placing the weight block composed of non-volatile memory (NVM) in a separate weight die using 3D integration technology and shared with pixels using modified rolling shutter operation. This advancement not only enhances the density and scalability of the pixel array but also opens up new possibilities for deploying more complex neural network models directly onto the hardware. Through detailed simulation results on TSMC 28nm, we demonstrate the effectiveness of our design in performing dot product operations— a fundamental building block of neural network computations— with improved efficiency and adaptability compared to existing designs. Furthermore, we present a novel machine learning framework compatible bucket-select curvefit approach to accurately model the non-linearity associated with analog convolution operation

% and showed with pixels using modified rolling shutter operation while maintaining computing parallelism
For this paper in section \ref{sec:bg} we discuss previous works that are related to in-pixel computing, section \ref{sec:circuit} introduces the novel NVM in-pixel computing circuit and the reconfigurability for all key ML parameters including weight value, kernel size, stride, channels, pixel skipping and demonstrate a novel bucket select curvefit approach to capture the non-linearity in the analog output voltage suitable for incorporation in standard ML training frameworks. Section \ref{sec:curvefit} introduces a new modeling method for FPCA circuit analog convolution output. Then section \ref{sec:sim} will elaborate on the simulation results and the analysis of the FPCA pixel circuit. The last section, section \ref{sec:con} concludes the  paper. 
\section{Background and Related Work}
\label{sec:bg}

In the first layer of convolutional neural networks (CNNs), the initial processing involves the multiplication of pixel outputs from the camera sensor with multi-bit weight values, a critical step in data interpretation and analysis \cite{algo_channel}. To facilitate this within the pixel array without compromising the resolution of the CIS, previous approaches \cite{datta2022processing,datta2022ace,kaiser2023technology} embed these weights directly into the pixel architecture. This embedding is made possible through the use of advanced 3D integration technologies, which allow for the vertical stacking of weights \cite{samsung_3D, sony_3D}. The physical implementation of these weights can be achieved through adjusting CMOS transistor geometry or by leveraging the resistance states in various non-volatile memory (NVM) devices such as Resistive Random Access Memory (RRAM), Phase Change Memory (PCM), and Magnetic Random Access Memory (MRAM) \cite{pip_mram}.

Additionally, the algorithm necessitates use of both positive and negative weight values to maintain accuracy in the test phase, requiring innovative circuit techniques to process and distinguish these weights accurately. In \cite{aps_p2m}, a novel use of the peripheral Single-Slope (SS) Analog-to-Digital Converter (ADC) is proposed. This approach combines the results of positive and negative weights by respectively increasing and decreasing the counter inside the ADC, enabling the calculation of the final convolution output. The integration of non-linear activation functions, such as the Rectified Linear Unit (ReLU), is ingeniously achieved by repurposing the on-chip Correlated Double Sampling (CDS) circuit found in CIS alongside the SS ADC. This setup ensures that the final ADC count, post CDS operation (which includes both ‘up’ and ‘down’ counting), results in a non-negative value, thereby effectively implementing the ReLU operation. Our proposed FPCA architecture also leverages this approach for the periphery ADC circuit design.

\begin{figure*}[!t]
\centering
\includegraphics[width = \textwidth]{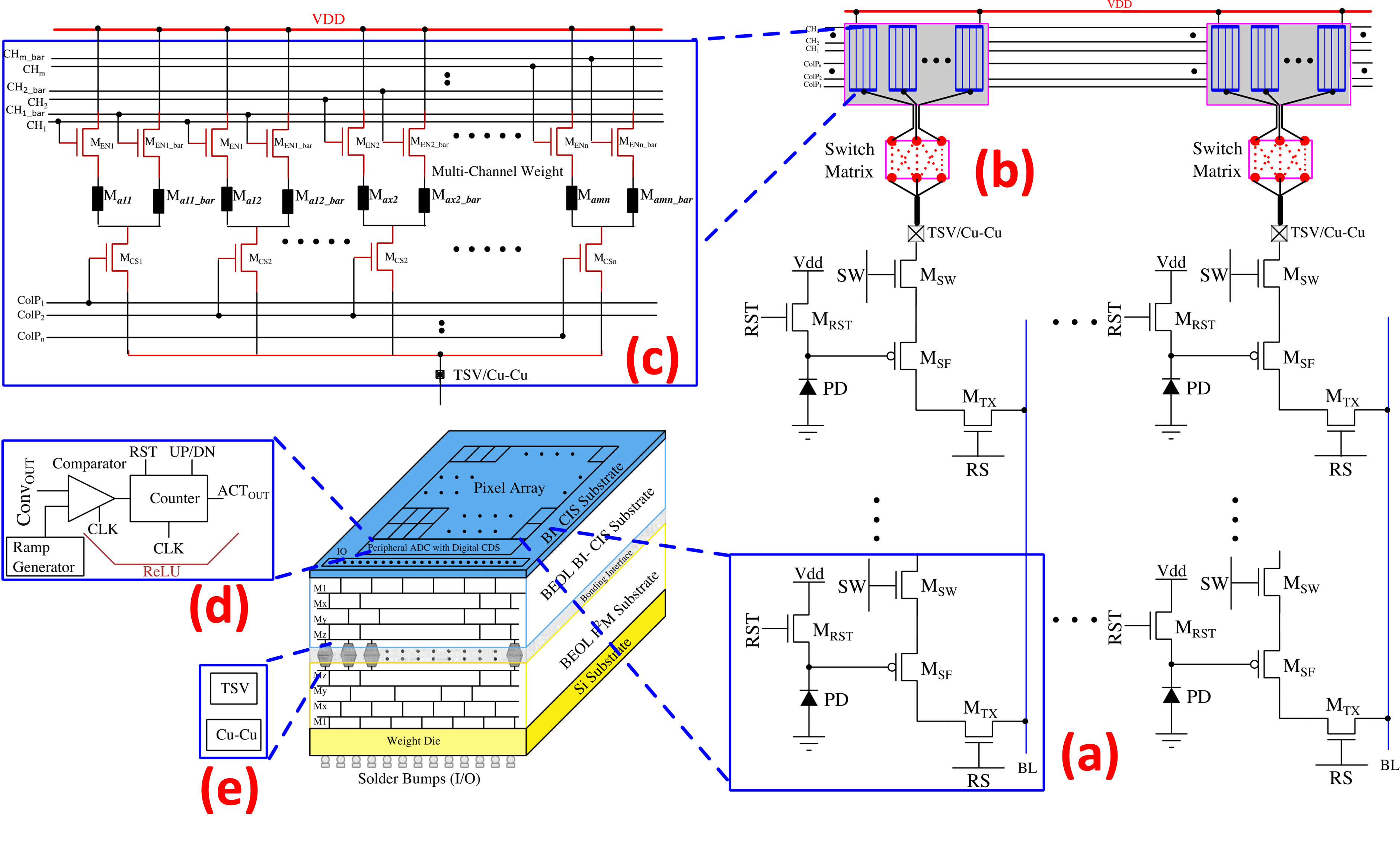}
\caption{Proposed circuit and the overall architecture of the FPCA, where (a) is the novel 4T APS schematic, (b) is the switch matrix that connects the pixel array to the shared weight block in a 3D integrated weight die, (c) is the example diagram of shared weight block, (d) is peripheral ADC and (e) is the connection between the two dies using either Through-Silicon Vias (TSV) or Copper-Copper bonding (Cu-Cu).}
\label{fig:whole_cir}
\vspace{-1mm}
\end{figure*}

Moreover, the integration of the batch normalization (BN) layer, crucial for training convergence, is partially combined with both the convolutional and ReLU layers. This is implemented by initializing the counter with the BN offset term and adjusting the weights with the BN scale term \cite{aps_p2m}. The processed activations are then ready to be transmitted off-chip via various I/O technologies such as low voltage differential signaling (LVDS), interposer (2.5D integration), through-silicon via (TSV), Cu-Cu bonding and Wireless, among others.\cite{yin2023design,kaiser2023technology}

\section{FPCA Architecture and Reconfigurability}
\label{sec:circuit}
% \subsection{Novel Pixel Circuit}
% \label{sec:circuit_architecture}

The essential innovation of the FPCA architecture is the reconfigurability it provides on different levels of the circuit. As shown in Fig. \ref{fig:whole_cir}, the proposed new circuit is composed of novel 4-transistor unit pixel circuit (Fig. \ref{fig:whole_cir}(a)), switch matrices (Fig .\ref{fig:whole_cir}(b)), multi-channel weight block for each pixel column (Fig. \ref{fig:whole_cir}(c)), and periphery ADC (Fig. \ref{fig:whole_cir}(d)). The pixel array and the multi-channel weight array are on separate dies that is connected by through-silicon vias (TSV) or Cu-Cu hybrid bonding using 3D integration (Fig. \ref{fig:whole_cir}(e))\cite{yin2023design}. These different blocks provide the architecture the reconfigurablity in all the key ML parameters: weight value, kernel size, channel size, stride size and the ability to achieve region skipping for neural network algorithms. For each column, it has one multi-channel weight block within the weight array that stores all the channels' kernel weight values using non-volatile memory (NVM). A weight value is represented by two NVMs ($M_{a11} \& M_{a11\_bar}$), one to represent the positive value and one for negative. If a weight value is negative then the NVM for positive value $M_{a11}$ stores 0 and the negative value is stored in the $M_{a11\_bar}$, the NVMs is connected to the source of transistor $M_{EN} \& M_{EN\_bar}$ that is used for selecting one channel in the output feature map. During a specific convolution operation, multiple pixels are activated and only one of the channels is enabled. Convolutions for multiple channels are performed sequentially. For each channel, the NVM for the positive weight is connected to the channel line $CH_i$ and the negative weight to $CH_{i\_bar}$, these two lines are activated sequentially, for positive \& negative signal accumulation. Further, each NVM storing a weight value is also connected to a kernel column select transistor $M_{CSi}$. The gate of these column select transistors are controlled by the input of Column Pattern control line $ColP$. The source of the $M_{CSi}s$ are then connected to the switch matrix. The output of the switch matrix would then connect to the drain of the transistor $M_{SW}$ within each unit pixel circuit. We will detail on these circuit structures and their interconnections in the following paragraphs.

\begin{figure*}[!t]
\centerline{\includegraphics[width=\textwidth]{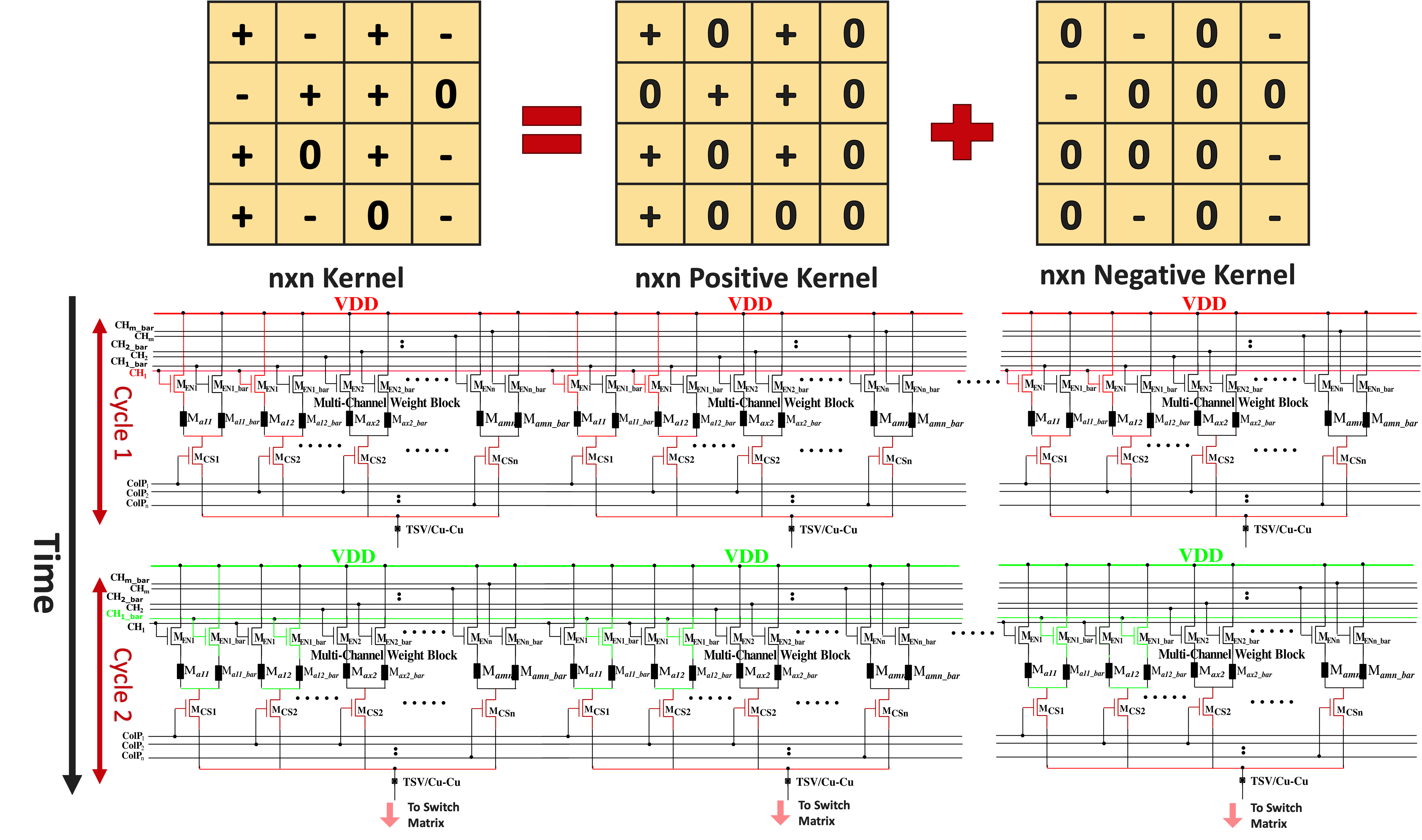}}
\caption{Detailed Multi-channel Weight Block (shared weight bock) Schematic with positive and negative kernel. The top part of the figure is the representation of the proposed method to store the kernel weight using one positive and one negative kernel, and the bottom part of the figure is the schematic representation of the two cycles of the multi-channel weight block to show the activated part of the circuit.}
\label{fig:weight_rep}
\end{figure*}

% \subsection{How does it work?}
\subsection{In-situ Multi-pixel Convolution Operation}

\begin{figure*}[!ht]
\centerline{\includegraphics[width=\textwidth]{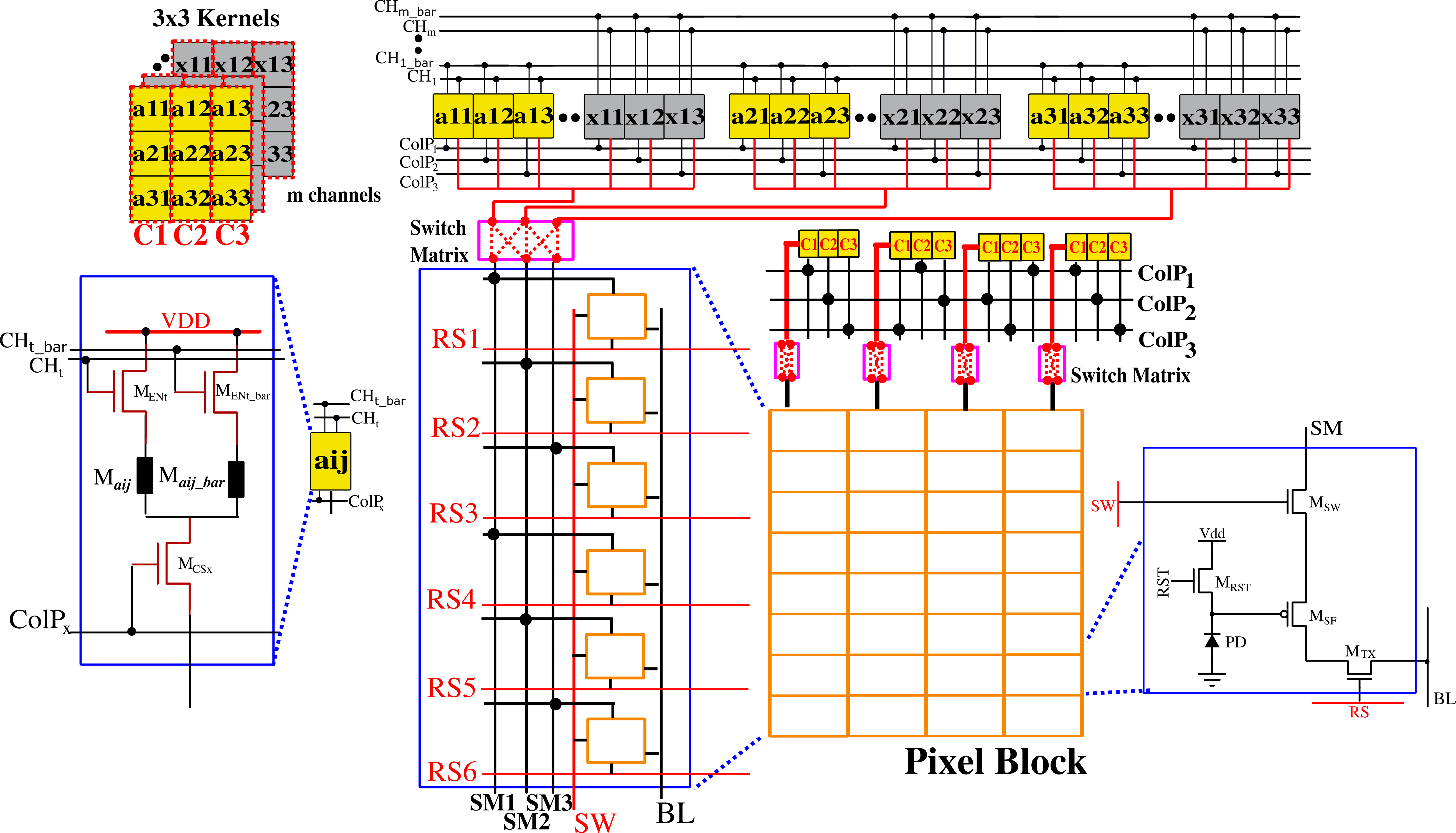}}
\caption{Detail architecture of the column design of FPCA pixel array and Multi-Channel Wight Block where (a) is pixel column design, (b) shows the connection pattern to different columns of the multi-channel weight block and the control signal $ColP$ (column pattern select line), (c) is example figure of a m-channel with max $3\times 3$ kernel and (d) is the pixel circuit design where the SW line is the column control line and RS is the row control line and the input to the pixel: line SM is connected to one node of the switch matrix.}
\label{fig:single_column}
\end{figure*}

To achieve the convolution operation, we simultaneously activate multiple pixels. For example, we activate $n_X{\times}n_Y{\times}3$ pixels at the same time, where $n_X$ and $n_Y$ denote the spatial dimensions and $3$ corresponds to the RGB (red, blue, green) channels in the input activation layer. For each activated pixels, the pixel output is modulated by the photo-diode current and the weight value of the activated $M_{a_{ij}}$ NVM in the multi-channel weight block shown as Fig. \ref{fig:whole_cir}(c) associated with the pixel. 
%The weight transistors are activated by respective select lines connected to their gates. 
% \rev{For a given convolution operation only one weight transistor is activated per pixel, corresponding to a specific channel in the first layer of the CNN. The weight transistors $W_i$ represent multi-bit weights through their driving strength. %When multiple pixels are activated, simultaneously, depending on light intensity and associated weight, the source follower transistors in respective pixels together pull up the voltage at the output node (labeled as \textit{Analog Convolution Output} in Fig. \ref{fig:pip_circuit}) to a higher level. 
In the FPCA architecture, the unit pixel circuit schematic follows up on previous works\cite{datta2022processing,kaiser2023technology,yin2023design}. For each pixel in the FPCA pixel array, the output voltage approximates the multiplication of light intensity and corresponding weight in the multi-channel weight array.
%is higher for higher light intensity as well as for higher width of weight transistors. 
For each output bit line, shown as vertical blue lines BL in Fig. \ref{fig:whole_cir}(a), the cumulative pull up strength of the activated pixels connected to that line drives it high.
The increase in pixel output voltages accumulate on the bit lines BL implementing an analog summation operation. Consequently, the voltage at the output of the bit lines represent the convolution operation between input activations and the stored weight in the weight die.

In summary, the presented unit pixel circuit can perform in-situ multi-bit, multi-channel analog convolution operation inside the FPCA pixel array, wherein the input activations are within the individual pixel (photodiode current) and the network weights are present in a shared weight block along with associated metal interconnects in the separate weight die connected to the pixel chip using TSV or Cu-Cu bonding\cite{kaiser2023technology,yin2023design}. 

\begin{figure*}[!t]
\centerline{\includegraphics[width=\textwidth]{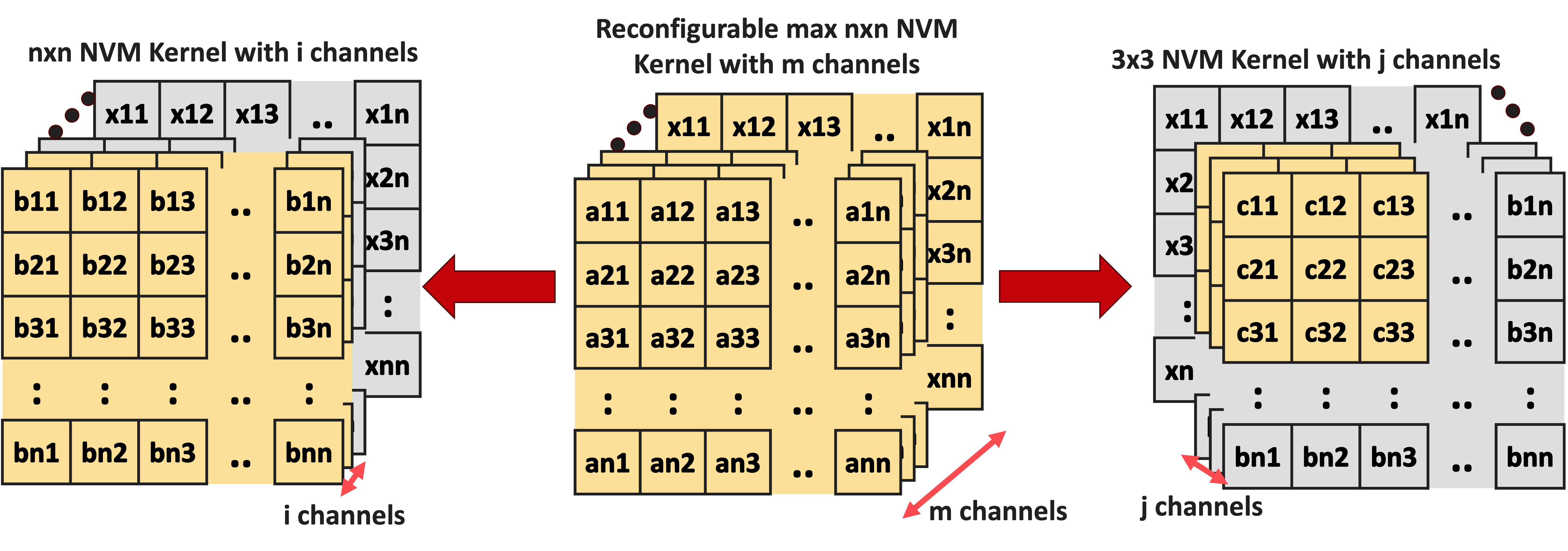}}
\caption{Reconfigurability in weight value, kernel size and channel size, the center of the figure is representing m channels of $k\times k$ kernels, the left part shows reconfigured weights in $i$ channels of $k\times k$ kernels and the right part of the figure shows the reconfigured $j$ channels for smaller $3\times 3$ kernels.}
\label{fig:reconfig_kernal}
\end{figure*}

\subsection{Shared-Weight Block for In-pixel Convolution Operation}
For the weight representation, as shown in Fig. \ref{fig:weight_rep}, a $n\times n$ kernel with both positive and negative weight values can be treated as two $n\times n$ kernels, one that stores only positive weight values and the other that stores the negative values. For negative weight values in the original kernel, a zero is stored in the corresponding position within the positive kernel. Conversely, where the original kernel has positive weights, zeros are placed in the corresponding positions of the negative kernel. This scheme requires a total of 2 cycles for 1 in-pixel convolution operation. As shown in the bottom part of Fig. \ref{fig:weight_rep}, for the first cycle, the $CH_i$ line would be pulled up to VDD to activate the positive weight values within the kernel (as shown by the red lines in Fig. \ref{fig:weight_rep}). And in the second cycle the $CH_{i\_bar}$ line would be pulled up while deactivating others to activate the negative weight through its corresponding $M_{EN\_bar}$ transistor, as shown by green lines in Fig. \ref{fig:weight_rep}. The activated NVMs in each cycle would then connect to the switch matrix and then to a certain pixel within the pixel array. Note that for a single channel, all positive weights' $M_{ENi}$ transistors' gate are connected to the same channel select line $CHi$, and all negative weights' $M_{ENi\_bar}$ transistors are connected to the complement channel select line $CHi\_bar$. 
% Therefore, in total if we have $m$ channels with maximum kernel size $n\times n\times 3$ ($3$ corresponds to the RGB (red, blue, green) channels in the input activation layer, note that in our system, the Red (R), Green (G), and Blue (B) channels can be concurrently activated, effectively unifying them as a singular operational unit. The input activation layer consists of three channels: red, green, and blue (RGB). 

Note, a typical RGB camera has 3 input channels. Our system allows these three color channels to operate concurrently by sharing channel select lines. If we have a total of $c_o$ output channels, the maximum kernel size across all output channels is $n \times n\times 3$, where the 3 corresponds to the RGB channels being processed together.
This simultaneous activation across the 3 input channels ensure that the RGB channels are processed simultaneously and their convolution outputs are accumulated together in accordance with the algorithmic requirement for modern CNNs. Consequently, the FPCA architecture would have $2\times c_o$ channel select lines ($CHi$s and $CHi\_bar$s), and each column of the pixel array would have in total $2\times n^2\times 3\times c_o$ number of weights per pixel column that is on a separate weight die. Note, this scheme allows the weights to be shared along columns, significantly reducing the number of transistors for representing weight kernels.
% Note, only one NVM weight output of every row of the $n\times n$ kernel are connected to a TSV/Cu-Cu at a given time since for each convolution operation, only one column of the kernel is mapped to one pixel column which means that for each row of the kernel, only one column of weight is being selected from a given weight bank column. In Fig. \ref{fig:weight_rep}, the circuit diagram demonstrates that for each grouping (the red line that connects to the $M_{CSi}$ transistors) only one NVM and one $M_{EN}$ transistor is being activated.

% \subsection{Pixel Array Functioning for In-pixel convolution operation of the FPCA}
\subsection{Mapping of Weights in Shared Weight Block with the Pixel Array for Convolution Operation}

As described earlier, Channel Select Line $CHi$ selects weight transistors $M_{CSi}$, similarly Column Pattern Select Line $ColPi$ and the switch matrix control the mapping of the weight block to a certain pixel column. Fig. \ref{fig:weight_rep} shows the detail schematic for the connection of the multi-channel weight block and the channel select transistors $M_{EN}$s, in addition Fig. \ref{fig:single_column} shows the detailed circuit connection of the rest of shared weight block and the pixel column including switch matrix.
As shown in Fig. \ref{fig:single_column}(c), for an example $3\times3$ kernel the different rows (row1: $a11, a12, a13$ etc., row2: $a21, a22, a23$ etc.) are connected to the different input nodes of the switch matrix, the $a11,a12,a13$ that belong to the first row are connected to the first red node of the switch matrix while $a21,a22,a23$ are all connected to the second node. The ColP lines (as shown in Fig. \ref{fig:single_column}(b)) are being used in order to select a specific column within a kernel for mapping the weight to the corresponding pixel column. For example, if ColP1 is being pulled up, C1 ($a11,a21,a31$) is connected to column 1 of pixel block, C2 ($a12,a22,a32$) is connected to column 2 of the pixel block. Each different ColP line enables assigning a different column of the kernel to a specific pixel column. Therefore, with the maximum kernel size of $n\times n$, there would be a total of $n$ lines of ColPs for column pattern selecting (as one pixel column could be assigned to every column of the weight kernel). This pattern ensures that any column of the kernel could be mapped to a specific pixel column.

The bottom $n$ nodes of the column switch matrix connect to the $n$ lines of SM that is connected to the transistor $M_{SW}$ within each pixel. For example, in Fig. \ref{fig:single_column}(a), there are 3 SM lines. Note that the SM lines are routed in 3D integrated chip and these lines would incur no area overhead for the pixel array. Each pixel would connect to one SM line. Pixels in row 1 connect to the $SM1$, pixel in row 2 connects to $SM2$ and row n connects to the $SMn$, then for the row n+1, it would connect to the $SM1$, every next row would follow the same pattern. Therefore, through this design, it is made sure that every nearby column-wise $k$ pixels (as the maximum kernel size would be $n\times n$) would be connected to a separate $SM$ line that links to a unique node within the switch matrix. The switch matrix is configured to route these $n
n$ separate $SM$ lines to correspond to the $n$ different weights in the specific column of the kernel.

As for the control signals, as shown in Fig. \ref{fig:single_column}(d), within each column pixel there are two control lines, shown in red, $RS$ and $SW$,  used for horizontal and vertical enabling the unit pixel. The $SW$ line is used for column enabling and the $RS$ line is used for row enabling. When both RS \& SW lines are enabled simultaneously the output of the corresponding unit pixel is connected for read operation. A more general diagram of the mechanism is shown in Fig. \ref{fig:single_column}(a). Each row shares the same $RS$ control line while every $SW$ line are shared among all pixel units within a column.

Therefore, for the overall matrix multiplication operation for each FPCA pixel column with $k\times k$ kernel:  channel select line $CHi$ and the column pattern select line $ColPi$ enables the specific $k$ weights in the $k\times k$ kernel to be connected to the n nodes in switch matrix; and the switch matrix controls the connecting pattern of the n SM lines that connect to the pixels in the pixel array. 
% Note, only one NVM weight for every row of the $n\times n$ kernel is connected to a TSV/Cu-Cu hybrid bond at a given time since for each convolution operation, only one column of the kernel is mapped to one pixel column which means that for each row of the kernel, only one column of weight is being selected from a given weight bank column. In Fig. \ref{fig:weight_rep}, the circuit diagram demonstrates that for each grouping (the red line that connects to the $M_{CSi}$ transistors) only one NVM and one $M_{EN}$ transistor is being activated.

\subsection{FPCA Reconfigurability}
Leveraging the circuit structures described above, the proposed FPCA architecture offers seamless reconfigurability on different levels, here we will discuss each level thoroughly.

\subsubsection{Reconfigurable Kernel Size}
\label{rec_kernel}
% On kernel size, here we will show it offers all kernel size that is small or equal to the maximum supported nxn kernel size.

\begin{figure*}[!th]
\centerline{\includegraphics[width=\textwidth]{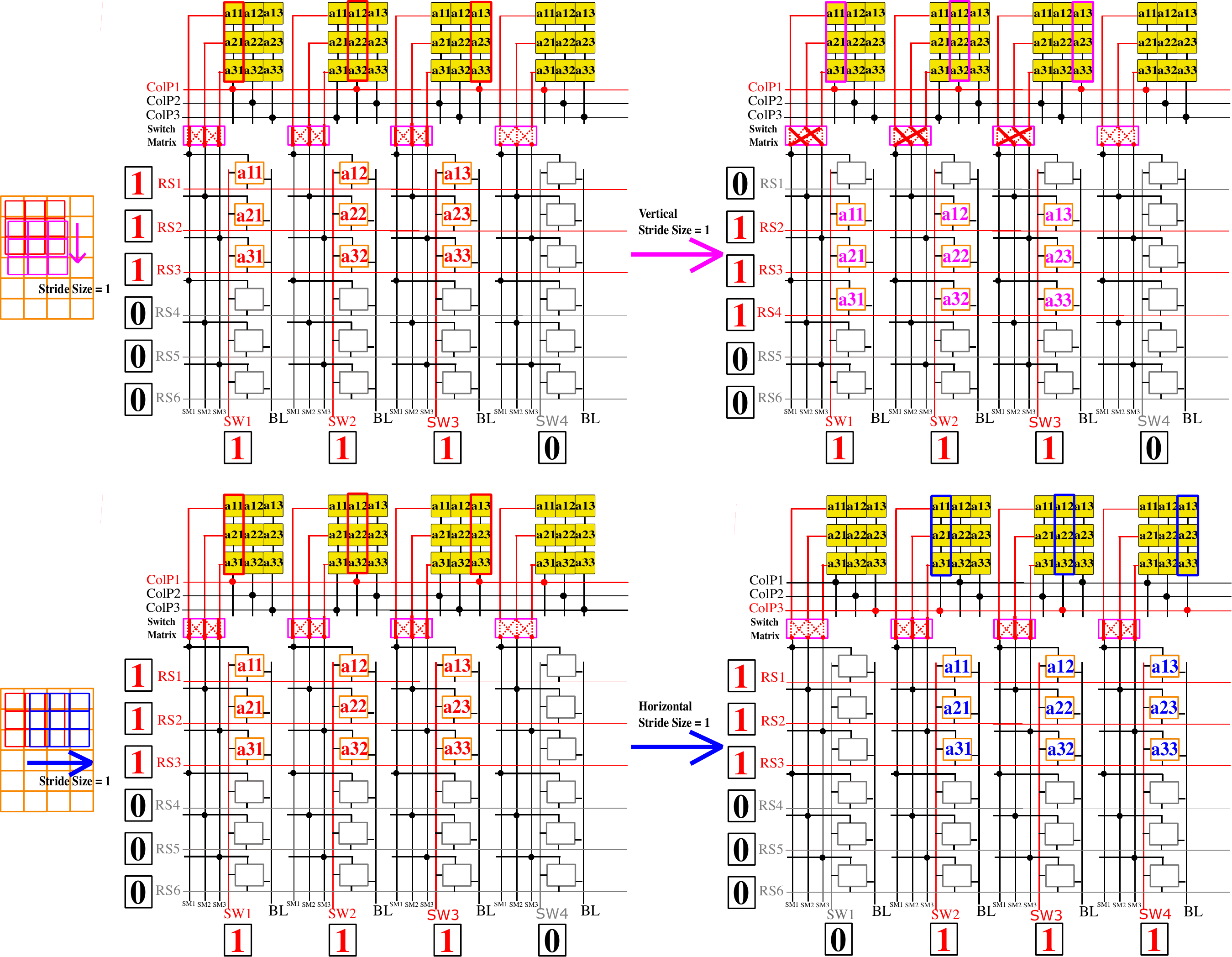}}
\caption{Figure representing vertical and horizontal striding of size $s = 1$ within the FPCA array.}
\label{fig:reconfig_stride}
\end{figure*}
% As there is no control signal that can directly deactivate parts of the kernel in the multi-channel weight block outside the pixel array, every time the whole kernel is being mapped into the pixel array. we need to apply other methods to achieve this functionality. There are two ways in our design to reconfigure the kernel size as shown in Fig. \ref{fig:reconfig_kernal}.
Given the absence of a direct control mechanism to selectively deactivate segments of the kernel within the shared weight block external to the pixel array, our system invariably maps the entire kernel into the pixel array for each convolution operation. To circumvent this limitation and to introduce kernel size reconfigurability, our design employs the strategy of writing 0 weight values to multi-channel weight block, as is illustrated in Figure \ref{fig:reconfig_kernal}:
% As we have prefixed the maximum kernel size ($n\times n$). any kernel size smaller than $n\times n$ can be achieved by writing 0 weight values to unoccupied slots in the maximum kernel. As the kernel weight values are pre-written, this method does not require additional time or control in execution. Note that by this method the pixel array's column output BL would always have the same number of activated pixels for each column (there will always be the maximum kernel size number of n pixels activated within each column).

In our design, we have established a predetermined maximum kernel size of $n\times n$ for each channel. To accommodate arbitrary kernel sizes smaller than the maximum, our approach involves assigning zero weight values to the slots not utilized within the maximum kernel configuration. This technique leverages the pre-loading of kernel weight values before the convolution operation for inference-only task, thus adding no overhead during compute operation. It is important to note that with this method, the output bit line (BL) of each column in the pixel array will consistently reflect the activation of a fixed number of pixels, corresponding to the maximum column size n for a kernel size of $n\times n$. This ensures uniform number of pixels are activated, irrespective of the actual kernel size that may vary based on specific CV application.
%     \item  Deactivate parts of pixel array: by controlling the RS line for row control and SW line for column control, we can deactivate parts of the kernel by giving a '0'
%     value to the required SW and RS lines as shown in Fig. \ref{fig:reconfig_stride}. Note that for this method, as the number of activated pixels within each column would vary, the output ADC that takes in the BL value for each column would need to readjust its range.
% \end{enumerate}
% \begin{figure}[!th]
% \centerline{\includegraphics[width=\linewidth]{Figs/Fig8_curvefit_bucket.png}}
% \caption{Diagram for the first two steps of the novel Curvefit Bucket Selection Function.}
% \label{fig:curvefit_buc}
% \end{figure}
% \begin{figure}[!th]
% \centerline{\includegraphics[width=\linewidth]{Figs/stepf.jpg}}
% \caption{Diagram for using Sigmoid Function to Replace Step Function.}
% \label{fig:stepf}
% \end{figure}

\begin{figure}[t!]
    \centering
    \begin{subfigure}[b]{\linewidth}
        \centering
        \includegraphics[width=\textwidth]{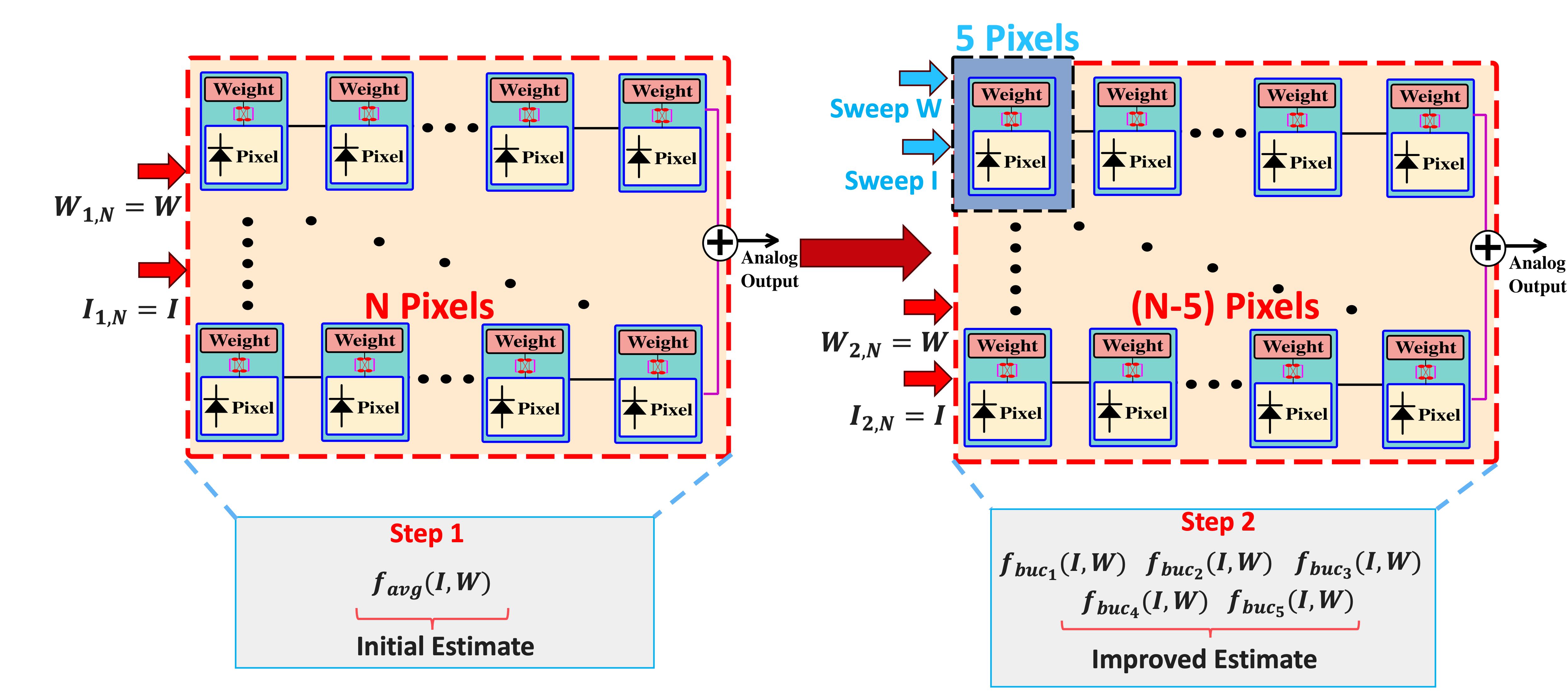}
        \caption{}
        \label{fig:curvefit_buc}
    \end{subfigure}%
    \vfill
    \begin{subfigure}[b]{\linewidth}
        \centering
        \includegraphics[width=\textwidth]{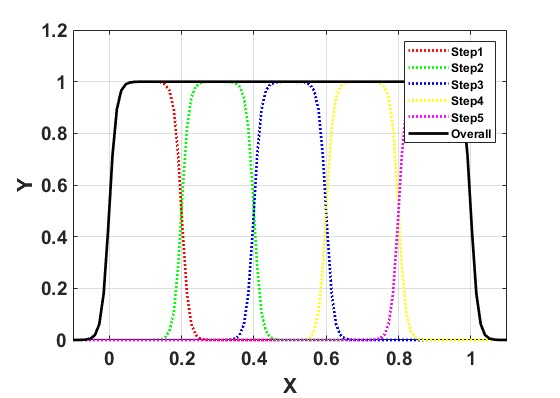}
        \caption{}
        \label{fig:stepf}
    \end{subfigure}
    \caption{Conceptual figure showing modeling approach of the proposed  bucket select curvefit function. (a) is the diagram for the two steps of the novel curvefit bucket selection function. (b) is the figure depicting use of sigmoid functions to replace step functions.}
\end{figure}

\begin{figure*}[!th]
\centerline{\includegraphics[width=\textwidth]{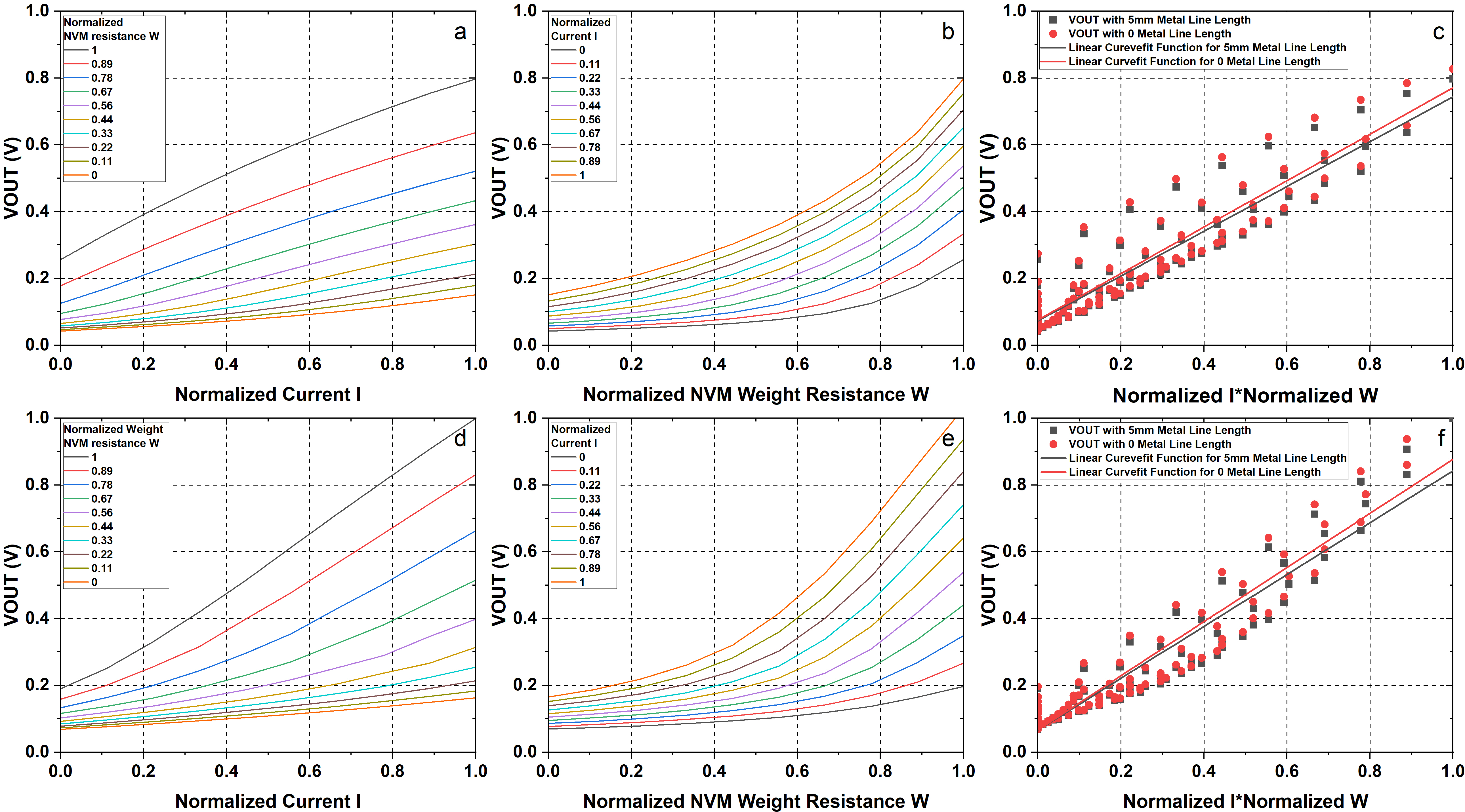}}
\caption{Simulation results of the FPCA circuit where (a) and (b) is the single pixel output v.s. normalized NVM weight resistance W and current I ( representing light intensity), in (a) each line corresponds to different resistance and in (b) each line corresponds to a specific value of input current. Scatter plot (c) is the linear curvefit function of the input data where the black and red color represents different metal line resistances between the weight die (shared weight block) and the pixel array. Plot (d), (e) and (f) are are similar to (a), (b) and (c), respectively but for 75 pixels (kernel size $5\times 5\times 3$) activated together performing convolution operation as opposed to a single pixel activation.}
\label{fig:sim_res}
\end{figure*}
\subsubsection{Reconfigurable Channel Size}
Shown in Fig. \ref{fig:reconfig_kernal}, as each column of the pixel array connects to the multi-channel shared weight block that stores weights for all the $c_o$ channels of maximum $n\times n$ kernel size, to reconfigure the channel size, we can simply control which CH line (channel select) line to activate as shown in Fig. \ref{fig:whole_cir}(c).
\subsubsection{Reconfigurable Stride Size}
Our proposed FPCA structure allows any stride size ranging from 1 to n, where n is the maximum kernel column or row number. Fig. \ref{fig:reconfig_stride} is the example schematic for the FPCA implementing both vertical stride and horizontal stride size of minimum size $s = 1$. For the operation, when striding vertically, the column pattern control signal ($ColP$) stays the same when the kernel moves down as the kernel is still mapping to the same pixel array columns. By rerouting using the switch matrix, the different $SM$ lines  can be reorganized to connect to different columns of the kernel. For example, as represented in Fig. \ref{fig:reconfig_stride}, switch matrix is rerouted  to allow vertical striding in upper half of the figure. The SW lines are further activated in accordance with the kernel size being mapped in the pixel arrays. As the kernel strides vertically, different $RS$ lines are pulled up in each cycle, while the $SW$ lines stay the same. 
For reconfiguring vertical stride size, the activation scheme for RS lines changes according to the desired stride size, while also reconfiguring the switching matrix. 

For horizontal stride, Fig. \ref{fig:reconfig_stride} shows the working schematic for horizontal stride size $s = 1$. When the kernel moves horizontally, the switch matrix routing would stay the same, the column pattern control line for ColPs  would be selected in accordance with the stride size. For example, for horizontal stride size of 1, $ColP1$ activation is followed by $ColP3$ activation. This ensures when $ColP1$ is active, the first column of the kernel is mapped to the first column of the pixel array, while when $ColP2$ is active the second column of the kernel is mapped to the first column of the pixel array, implementing the horizontal striding. Meanwhile to select different columns different groups of the $SW$ lines would be ON. 

This configuration is essential for allowing the weight kernel stride horizontally in the pixel array, the total of n $ColP$ lines (maximum kernel size is $n\times n$) allows the kernel to move to every location in the pixel array horizontally. Note that Fig. \ref{fig:reconfig_stride} represents single kernel convolution operation, the FPCA structure allows massive parallel convolution operations along a set of selected rows in the pixel array.
\subsubsection{Multi-Cycle Convolutional Operations}
As mentioned in section \ref{rec_kernel}, for any kernel size that is smaller than the maximum $n\times n$, 0 weight values would be written into the NVMs that are not occupied in the predetermined maximum kernel for each channel. Therefore, kernels of size $n\times n$ are applied to the pixel array, although some weights are set to hold 0 value. But as the stride size vary, it would require different number of cycles to complete the convolution operation for all stride locations. For a kernel with $n$ for the maximum kernel width of one channel with a depth of 1, the total cycle needed for a stride size of $S$ is $\frac{lcm[S,n]}{S}$. Thus, the total cycle number $N_C$ needed for generating the output kernels for the next layer of convolution would be
\begin{equation}
    N_C = 2\times h_o \times c_o\times \frac{lcm[S,n]}{S} \label{n_C}
\end{equation}
where $h_o$ is the height of the output kernel, $c_o$ is the number of output channels, $S$ is the stride size, $n$ is the maximum kernel width. 
\subsubsection{Pixel Region Skipping}
The FPCA architecture incorporates the region skipping functionality, achieved by the management of row-wise (RS) and column-wise (SW) control lines as depicted in Fig.  \ref{fig:reconfig_stride}. To enable unit-wise control over which specific rows and columns of the pixel array are activated, a total of $R_P\times C_P$ number of SRAMs along the periphery of the pixel array circuit would be needed. Here, $R_P$ and $C_P$ represent the total number of rows and columns in the pixel array, respectively. These SRAMs are intended to store control data for the RS and SW lines. However, $R_P\times C_P$ number of SRAMs is considerably high. Such an approach would result in substantial area overhead. Consequently, a block-wise approach to region skipping is recommended. This method reduces the complexity and number of SRAMs needed by grouping pixels into blocks and storing a single set of RS and SW values for each block. This strategy not only minimizes the area overhead but also simplifies the control scheme, making it a more practical solution for efficiently managing region skipping within the FPCA framework. For example, if the entire pixel array is divided into 8x8 blocks, then only 64 SRAM locations are needed to hold the region skipping data.
\section{Accurate Modeling of Analog Convolution  through Bucket-Select Curvefit Function}
\label{sec:curvefit}
Analog computing including the proposed FPCA system shows inherent non-linearity due to non-linear behavior of constituent devices including NVM and transistors. This non-linear behavior needs to be accurately modeled in the algorithmic framework to mitigate any accuracy loss \cite{aps_p2m_detrack,bdd100k}. 
However, a significant challenge arises as each pixel's output  is influenced by the cumulative operation of other pixels that are connected together and activated simultaneously to perform parallel dot product operation. This inter-dependence of a given pixel's behavior on other pixels complicates accurate and machine learning framework compatible modeling of analog computing performed in the FPCA. 
%As the in-pixel computing enabled pixel array gives the input for imaging algorithms that cooperates with machine learning. Traditional simulation tools such as HSpice while robust for circuit analyses and can output accurate result, fail to integrate effectively with backend machine learning algorithms. As machine learning algorithms can learn from the input non-linearity and improve its effectiveness, it is crucial for the FPCA design to develop an innovative algorithm that employs an equation-based method to model the non-linearities directly. This approach not only bypasses the limitations of HSpice but also enables the algorithm to learn from and adapt to the unique characteristics of each pixel output.

%\subsection{Bucket Select Curvefit Approach}

We propose a novel two step bucket-select curvefit method to accurately model non-linear behavior associated with the analog computing FPCA system.
%Due to the high error rate in predicting the output of in-pixel convolution using only the curvefit function derived from sweeping the input light intensity I (photodiode current) and weight W (NVM resistance) while all the  pixels share the same value, the novel bucket selection curvefit approach is being applied. 
Fig. \ref{fig:curvefit_buc} shows the conceptual diagram representing step 1 and 2 of the proposed bucket select curvefit method. 
%The input to the model is a set of input currents $I_i$s and Weights $W_i$s. 
In step 1, an initial estimate for the analog output voltage is obtained by using a \textit{generic} analytical function. This generic analytical function is obtained by curve fitting a 2D surface plot to the analog SPICE output of N pixels, where input current to all the N pixels are kept the same and swept within a range of minimum and maximum current values. Similarly, the weights associated with all the N pixels are also kept the same, and swept within a range of minimum and maximum weight values. The analog output estimate obtained from this step is used as an input to step 2. Thus, a total of 2N parameters (N input current and N weights) along with the initial estimate of analog voltage serves as input to step 2.

The initial estimate categorizes the output into one of several predetermined smaller ranges, referred to as `buckets'. Each bucket, representing a segment of the total output range, is associated with a unique curvefit function. Thus, for each range the analog behavior is modeled by a curvefit function specifically tailored to the range of interest as determined by the initial estimate. For example, the overall analog voltage range from 0V to 1V can be sub-divided into 5 different buckets each of 200mV size. The initial estimate obtained in step 1 helps select the bucket (or range) of interest. For each bucket, a specific curvefit function models the behavior of the pixels (or the error associated with the initial estimate) accurately in the vicinity of the voltage range associated with the bucket. The analog output voltage predicted by step 2 is therefore, much more accurate compared to the initial estimate. This is because although the behavior of any pixel is dependent on the cumulative effect of other pixels it is connected to, yet the pixel's response is a strong function of its own input current and weight and only a weak function of the cumulative effect of other pixels. Thus, step 1 helps to estimate the cumulative effect of the pixels using a generic curvefit function, while step 2 uses a tailored curvefit function to accurately model the strong dependence of a pixel on its input current and its weight based on the initial estimate obtained from step 1. 

The bucket curvefit functions are obtained by modifying the simulation setup of step 1. Majority of the pixels still share same value of input currents and weights. The value is chosen such that the output is forced to be within the range of a specific bucket, mimicking the cumulative effect of the interconnected pixels. A small subset of pixels then undergo a parameter sweep, leading to the generation of a distinctive curvefit function specific to the bucket of interest.

%\textcolor{blue}{This new approach takes into consideration that each pixel in a function of the total pixel block output. In the first step, a general pixel output function is applied uniformly across each pixel's input to derive an average output, serving as the preliminary estimate. This estimate categorizes the output into one of several predetermined smaller ranges, referred to as 'buckets'. Each bucket, representing a segment of the total output range, is associated with a unique curvefit function. This segmentation into buckets allows for tailored modeling of the output. As for generating these bucket curvefit functions: initially, a uniform set of parameters is applied across all pixels, determining a baseline voltage output. Subsequently, the majority of the pixels are maintained at this standard parameter set, which is treated as a constant. A subset of pixels then undergoes a parameter sweep, leading to the generation of a distinctive curvefit function. This function is then evaluated through random testing to assess its error rate against actual simulation results. If the error rate remains below 5\% within a specified range of output voltages, the curvefit function is designated as the 'bucket curvefit function' for that voltage range. This method ensures that each bucket curvefit function accurately represents a specific segment of output voltages, enhancing the overall modeling precision and efficiency of the pixel array. 

For our simulations, we considered a kernel size of $5\times 5\times 3$, the overall model consists of  a total of 6 curvefit functions: the first one is the generic function: $f_{avg}$, it is obtained by sweeping I (Input current or light intensity) and W (Weight) wherein all 75 pixels share the same parameter values for input current and weights. The rest 5 are the bucket curvefit functions $f_{buc_1}$ to $f_{buc_5}$, they are generated by sweeping a small subset of 5 pixels' I and W (these 5 pixels share the same parameter) while keeping the the other 70 pixels' I and W values same ($I_{C},W_{C}$) and chosen so as to force the output in the specific range for a particular bucket of choice. 

\begin{figure*}[!ht]
\centerline{\includegraphics[width=\textwidth]{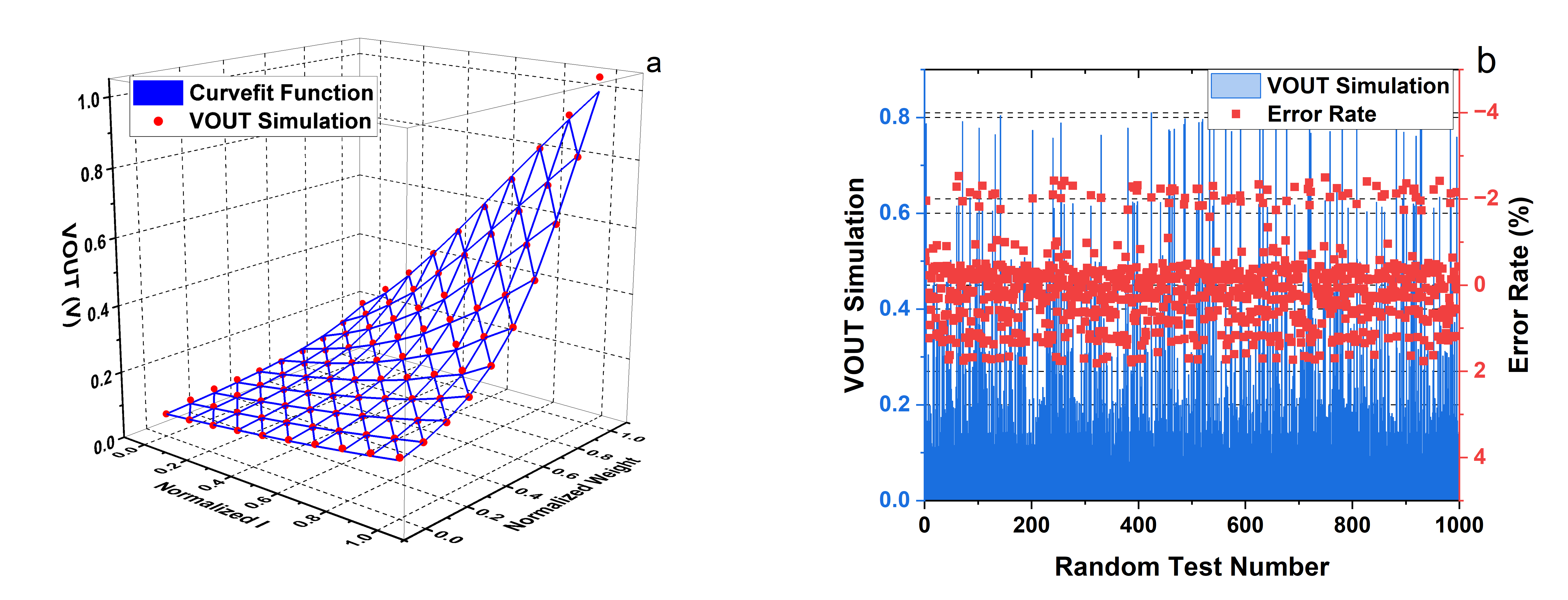}}
\caption{3D curvefit function plot and the error rate bar plot. Plot (a) is the 3D curvefit function used to model the analog output behavior of the circuit for algorithmic use, plot (b) shows the error rate after applying the bucket select curvefit function with respect to the simulation data using random inputs (input current and weights) to each of the 75 pixels in TSMC 28nm HPC+ technology.}
\label{fig:curvefit}
\end{figure*}

% \begin{equation*}
%     f_{buc_i}: f_{avg}(I_{C_i},W_{C_i})=V_{OUT_{max}}\times(\frac{i-1}{5}+\frac{1}{10}) (i\in  [1,5]  )
% \end{equation*}
With these 6 curvefit functions, our approach to monitor the circuit output with the total of 150 different parameters of I and W ($I_{0}-I_{74},W_{0}-W_{74}$) has three steps:
\begin{enumerate}
    \item Select 1 $f_{buc_s}$ out of the 5 bucket $f_{buc_1}-f_{buc_5}$ from the result $V_{OUT_{est}}$ from step 1, each bucket function $f_{buc_i}$ covers the range $[\frac{i-1}{5},\frac{i}{5}], (i\in[1,5])$.
    \item Use the selected bucket curvefit function to calculate the final predicted convolution output. Here the bucket curvefit function adjusts the initial estimate to obtain more accurate estimate for behavior of each pixel:
    \begin{equation*}
    \begin{split}
        V_{OUT_{pd}}=&\sum_{i=0}^{74}\frac{f_{buc_s}(I_i,W_i)-f_{avg}(I_{C_s},W_{C_s})}{5}\\
        &+f_{avg}(I_{C_s},W_{C_s})
    \end{split}
    \end{equation*}
\end{enumerate}
   This approach shows much more reliable output value as shown in the error rate bar plot Fig. \ref{fig:curvefit}(b), the error rate is below 3\%.\\
  To simplify the method, we combine step 1 and 2 into a single analytical equation that can be easily incorporated into ML frameworks like PyTorch. Selecting one out of five bucket curvefit functions effectively can be performed using a set of step functions. A significant drawback of using a step function is its non-differentiability at certain points, which poses a challenge for machine learning algorithms that require continuous derivatives for back propagation algorithm during training. To address this issue, we use sigmoid function ($\sigma(x)$) as an alternative to step function. The sigmoid function is advantageous because it is differentiable at all points, ensuring smoother transitions between output ranges. By combining sigmoid functions, as shown in Fig. \ref{fig:stepf}, the bucket selection process based $V_{OUT_{est}}$ can be accommodated into a single analytical equation: 
\begin{equation*}
\begin{split}
    V_{OUT_{pd\sigma}}= &\sum_{i=1}^{5}\biggl( \Bigl(\sigma(100(x-\frac{i-1}{5}))+\sigma(100(\frac{i}{5}-x))-1\Bigl)\\
    &\times\Bigl((\sum_{j=0}^{74}\frac{f_{buc_i}(I_j,W_j)-f_{avg}(I_{C_i},W_{C_i}}{5})\\
    &+f_{avg}(I_{C_i},W_{C_i})\Bigl) \biggl)
\end{split}
\end{equation*}

\section{Results and Discussion}
\label{sec:sim}

Fig. \ref{fig:sim_res} (a)-(c) and (d)-(f) show the simulation results for the analog convolution output for a single FPCA pixel and a group of 75 pixels, respectively, using TSMC 28nm HPC+ technology.  The figures also show scatter plot comparing the linearity of the circuit output with that of ideal mathematical convolution output. As seen in Fig. \ref{fig:sim_res}
(c) and (f), the simulated convolution output of the FPCA pixel circuit exhibits fairly linear behavior. The small non-linearity associated with the analog convolution output needs to be accounted for in a machine learning framework compatible model such that the model can be used for algorithmic training of a deep learning network to mitigate any accuracy loss.
Towards that end, the effectiveness of the novel curvefit bucket selection function approach in predicting circuit output is demonstrated in Fig. \ref{fig:sim_res}(b). This figure illustrates the error rate of estimated output voltage when applying the proposed curvefit method. The method involves simulation of convolution operation for 75 connected pixels with 150 randomly selected parameters ($Ws, Is$) spanning the entire range of weight and input current parameters. The comparison with simulation data shows that this method achieves accurate prediction of the analog output voltage, with an error rate of less than 3\%. Additionally, Fig. \ref{fig:sim_res}(c) and (f) display the simulated pixel output variations resulting from different metal length resistances between the shared-weight block and the unit pixel. With a metal line distance ranging from 0mm to 5mm, the difference in output voltage is minor. Therefore the presented curvefit model could be used for FPCA output modeling spanning a large range of metal length (0-5mm) between the unit pixel and the shared-weight block located in the weight die.

\begin{figure*}[ht]
    \centerline{\includegraphics[width=\textwidth]{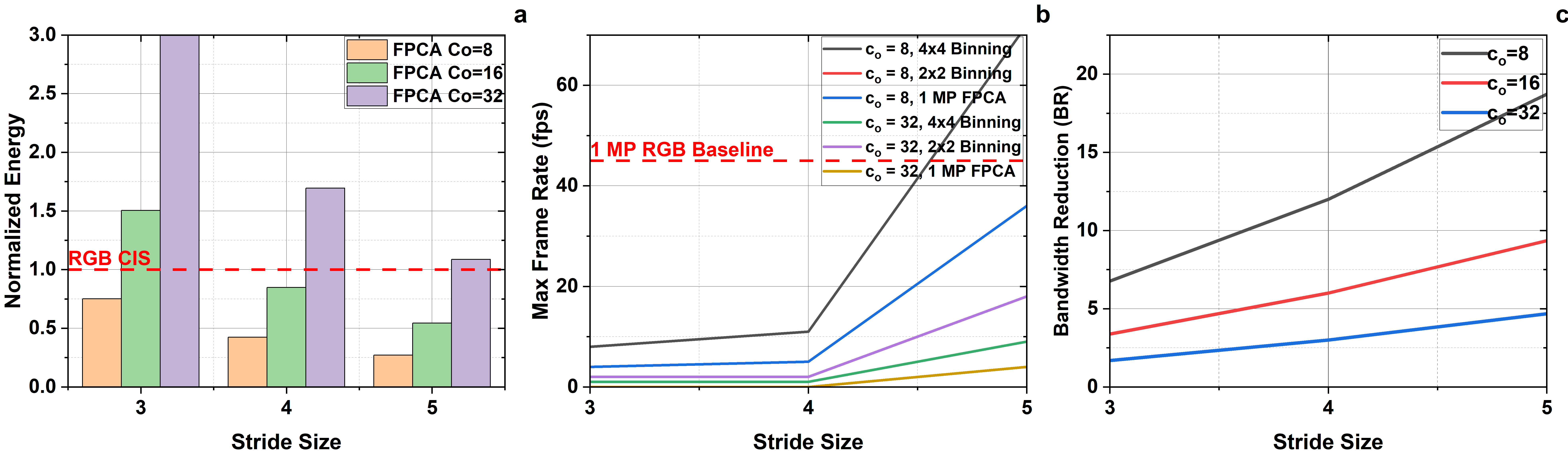}}
    \caption{(a) Energy analysis showing normalized energy versus stride size for different numbers of output channels considering kernel size $n\times n = 5\time 5$, (b) Latency analysis: Maximum frame rate versus stride size with different numbers of output channels and pixel binning considering the same kernel size of $5\times 5$, and (c) is bandwidth reduction (BR) analysis where BR versus stride size for different numbers of output channels is plotted.}
    \label{fig:3fig}
\end{figure*}

\subsubsection{Energy Analysis}
The total frontend energy consumption $E_{FRONTEND}$ is quantified using Eq. \ref{energy}, where $e_{PX}=148 pJ$(calculated from simulation result), $e_{ADC}=41.9 pJ$ \cite{kaiser2023technology}, and $E_{IO}$ represent the energy consumed per convolution operation, the energy per ADC read operation, and the total communication (IO) energy, respectively. The IO energy is further detailed in Eq. \ref{io_energy}, where $e_{IO}=12.34 pJ/bit$ indicates the energy cost per bit for the employed IO technology, specifically LVDS in this instance \cite{lvds_io}, $b_{ADC}=8$ represents the ADC bit precision, $h_o$, $w_o$, and $c_o$ specify the height, width, and number of output channels of the output activation map, respectively.  Fig. \ref{fig:3fig}(a) presents  the normalized energy consumption for various stride sizes and number of channels in the output feature map for a constant kernel size of $5\times 5$. The graph shows that employing strides of size 5 (non-overlapping) leads to maximum energy savings. The energy savings decrease as the stride size decreases, since lower stride size implies more number of convolution operation. Further, smaller number of output channels lead to higher energy savings. For reference, baseline energy number for a RGB camera without FPCA computing is shown in figure using a red dotted line. Thus, energy savings is achieved through use of higher stride size and lower number of channels \cite{chowdhery2019visual,bdd100k}. In contrast, increasing the output channel count to 32 does not lead to energy savings. This lack of improvement is due to an increase in the number of convolutional operations ($N_{C}$ as shown in Eq. \ref{n_C}). These findings highlight that for FPCA design and in general for analog convolution in pixel, algorithm optimizations to reduce number of channels and increase number of strides is necessary along with hardware design. Thus, algorithm-hardware co-design is an imperative aspect of analog computing in pixel as well as FPCA design. 
\begin{align}
    E_{FRONTEND} & = N_{C} \times (e_{PX} + e_{ADC}) + E_{IO}  \label{energy} \\
    E_{IO} & = h_o \times w_o \times c_o \times b_{ADC} \times e_{IO} \label{io_energy}
\end{align}
% \begin{figure}[!th]
% \centerline{\includegraphics[width=\linewidth]{Figs/Enerhy Plot.png}}
% \caption{Frontend energy trade-off analysis where the left is the normalized energy versus stride for different numbers of output channels considering LVDS IO, and the right is normalized energy versus stride for different IO configurations.}
% \label{fig:3fig}
% \end{figure}
\subsubsection{Latency Analysis}
The latency $T_{FRONTEND}$ of FPCA convolution operation, depends on the number of read cycles as dictated by Eq.\ref{n_C}. This latency, is further quantified using Equation \ref{Td}, where $T_{EXP}$ represents the exposure time, $T_{ADC}$ the ADC read time, and $T_{IO}$ the communication delay associated with IO pins. The IO delay, $T_{IO}$, is influenced by the ADC bit precision ($b_{ADC}$), the width of output activation map $w_o$, IO bandwidth ($BW_{IO}=1Gbps$ \cite{lvds_io}), and the total number of IO pads ($n_{IO(PAD)}=24$) utilized on the chip as in Eq. \ref{Tio}. Fig. \ref{fig:3fig}(b) shows the maximum frame rate ($\frac{1}{T_{FRONTEND}}$) achievable with an FPCA-enabled CIS, varying by stride number across different output channels and pixel binning scenarios, assuming a kernel size of 5. The plot shows that the maximum frontend frame rate of the FPCA model is generally lower than that of conventional RGB CIS. This is due to the inclusion of both positive and negative weights in the first convolutional layer and the fixed max kernel size for shared weight block. Nevertheless, by optimizing the stride, reducing output channels, and implementing pixel-level binning, higher frame rates are attainable (as demonstrated with $c_o = 8$ and $4 \times 4$ binning for stride number = 5).
\begin{align}
    T_{FRONTEND} & = N_{C} \times (T_{EXP} + T_{ADC} + T_{IO}) \label{Td} \\
T_{IO} & = \frac{w_o \times b_{ADC}}{BW_{IO} \times n_{IO(PAD)}} \label{Tio}
\end{align}
It is important to note that for in-pixel computing, as defined in \cite{kaiser2023technology} the frontend latency is considered as  the total computation time for all activations of the output feature map within the first convolutional layer. To reduce the latency compared to conventional CIS significant reductions need to be made in the output feature map dimensions, such as through fewer channels, larger strides, or binning.\cite{datta2022processing,9939582} Moreover, reduction in exposure time can also increase the frame rate, irrespective of the specific FPCA configuration employed.

\subsubsection{Bandwidth Reduction Analysis}
For Bandwidth Reduction in FPCA, the data bandwidth reduction (BR) can be estimated using Equation \ref{BR}\cite{kaiser2023technology}. Here, $I$ and $O$ represent the number of elements in the input RGB image and the output activation map of the first convolutional layer, respectively. The input $I$ is calculated as $h_i \times w_i \times 3$, where $h_i$ and $w_i$ are the height and width of the input image, and the factor 3 is for the three RGB channels. The output $O$ is derived from Equation \ref{O} and Equation \ref{ho_wo}, where $h_o$, $w_o$, and $p$ denote the height and width of the output activation map and the padding, respectively. The factor $\frac{4}{3}$ in Equation \ref{BR} accounts for the compression efficiency from converting the Bayer RGGB pattern, commonly used in image sensors, to the standard RGB format. And the term $\frac{12}{b_{ADC}}$ reflects the conversion from a higher bit depth, typically 12 bits per color channel in raw sensor outputs, to the bit precision ($b_{ADC}=8$ ) used by the peripheral ADC, which is set to 8 bit activations for deep learning applications.

\begin{align}
    BR & = \left(\frac{I}{O}\right) \left(\frac{4}{3}\right) \left(\frac{12}{b_{ADC}}\right) \label{BR} \\
    O & =  h_o \times w_o \times c_o \label{O} \\
    h_o (w_o) & = \frac{h_i(w_i) - n + 2\times p}{S} + 1 \label{ho_wo} 
\end{align}
Fig. \ref{fig:3fig}(c) illustrates the estimated data bandwidth reduction (BR) versus stride size for a kernel size of $5 \times 5$ with various numbers of output channels. The graph shows that the FPCA approach can provide a significant reduction in data bandwidth compared to conventional CIS, particularly when employing large strides and fewer output channels. Additionally, improved BR values can be achieved by implementing pooling (either max or average) at the periphery following the output from the first convolution layer in the FPCA design.

\section{Conclusion}
\label{sec:con}
In conclusion, this paper has introduced a Field-Programmable Pixel Array (FPCA), a novel approach to in-sensor and in-pixel computing that pushes the potential of pixel arrays within CMOS image sensors to wide range of modern convolutional deep learning networks. 
Unlike traditional pixel arrays, FPCA provides dynamic adaptability in weight values, kernel sizes, channel sizes, and stride sizes, addressing the rigid constraints that have limited intelligent sensor design and functionality without sacrificing pixel area by implementing weight array on a separate die connected with TSV or Cu-Cu bonding. 
Moreover, the FPCA architecture integrates the capability of region skipping, allowing for selective processing that enhances efficiency and reduces unnecessary computations. This functionality augments data processing by focusing on regions of interest while also significantly conserving energy and bandwidth, which are critical resources especially for extreme-edge applications. The paper also presents a novel bucket select curvefit based analog modeling approach that can accurately capture the non-linear behavior of analog computing. The resultant model is machine learning framework compatible and can be used for training deep learning networks to mitigate accuracy loss. Advantageously, the developed model is applicable to analog computing in general beyond the presented FPCA use-case, including memristive crossbar arrays.
Further, our analysis shows the dependence of energy, latency and bandwidth metrics on typical deep learning parameters, highlight the importance of algorithm-hardware co-design for extreme-edge intelligence applications.\\
In summary, the work provides a versatile framework with potential applications across wide range of various computer vision tasks, spanning from simple object detection to more complex scene understanding. Its flexibility and efficiency may contribute to the development of intelligent devices designed to perform sophisticated image processing tasks directly at the point of data generation.
\section{Acknowledgement}
 This work is supported in part by National Science Foundation under award CCF2319617.

\bibliographystyle{unsrtnat}
\bibliography{reference}  %%% Uncomment this line and comment out the ``thebibliography'' section below to use the external .bib file (using bibtex) .





% \end{thebibliography}

\end{document}